\documentclass[10pt,twocolumn,letterpaper]{article}

\usepackage{iccv}      
\usepackage{times}
\usepackage{epsfig}
\usepackage{graphicx}
\usepackage{amsmath}
\usepackage{amssymb}
\usepackage{csvsimple}
\usepackage{tabularx}
\usepackage{booktabs}
\usepackage{caption}

\definecolor{iccvblue}{rgb}{0.21,0.49,0.74}
\usepackage[pagebackref,breaklinks,colorlinks,allcolors=iccvblue]{hyperref}

\def\paperID{11025} 
\def\confName{ICCV}
\def\confYear{2025}

\title{CAD-Coder:Text-Guided CAD Files Code Generation}

\author{Changqi He\\
{\tt\small hechangqi235@outlook.com}
\and
Shuhan Zhang\\
{\tt\small tigerzh7@hrbeu.edu.cn}
\and
Liguo Zhang\\
{\tt\small zhangliguo@hrbeu.edu.cn}
\and
Jiajun Miao\\
{\tt\small miaojiajun@hrbeu.edu.cn}\\
College of Computer Science and Technology, Harbin Engineering University, China
}

\begin{document}
\maketitle
\begin{abstract}
Computer-aided design (CAD) is a way to digitally create 2D drawings and 3D models of real-world products. Traditional CAD typically relies on hand-drawing by experts or modifications of existing library files, which doesn't allow for rapid personalization. With the emergence of generative artificial intelligence, convenient and efficient personalized CAD generation has become possible. However, existing generative methods typically produce outputs that lack interactive editability and geometric annotations, limiting their practical applications in manufacturing. To enable interactive generative CAD, we propose CAD-Coder, a framework that transforms natural language instructions into CAD script codes, which can be executed in Python environments to generate human-editable CAD files (.Dxf). To facilitate the generation of editable CAD sketches with annotation information, we construct a comprehensive dataset comprising 29,130 Dxf files with their corresponding script codes, where each sketch preserves both editability and geometric annotations. We evaluate CAD-Coder on various 2D/3D CAD generation tasks against existing methods, demonstrating superior interactive capabilities while uniquely providing editable sketches with geometric annotations.
\end{abstract}






    
\section{Introduction}
\label{sec:intro}

Generative AI is profoundly reshaping the production mode and innovation path in  industrial field, and promoting the intelligent transformation of traditional methods in machinery,construction, automobile and other fields. Computer-aided design (CAD) holds significant importance in industrial design. However, traditional CAD models require manual drafting by professionals, which demands a certain level of expertise from engineers, and is often inefficient and prone to errors. In recent years, with the advancement of generative AI and 3D generation, research related to CAD generation has garnered widespread attention.

\begin{figure}[t]
  \centering
    \includegraphics[width=\linewidth]{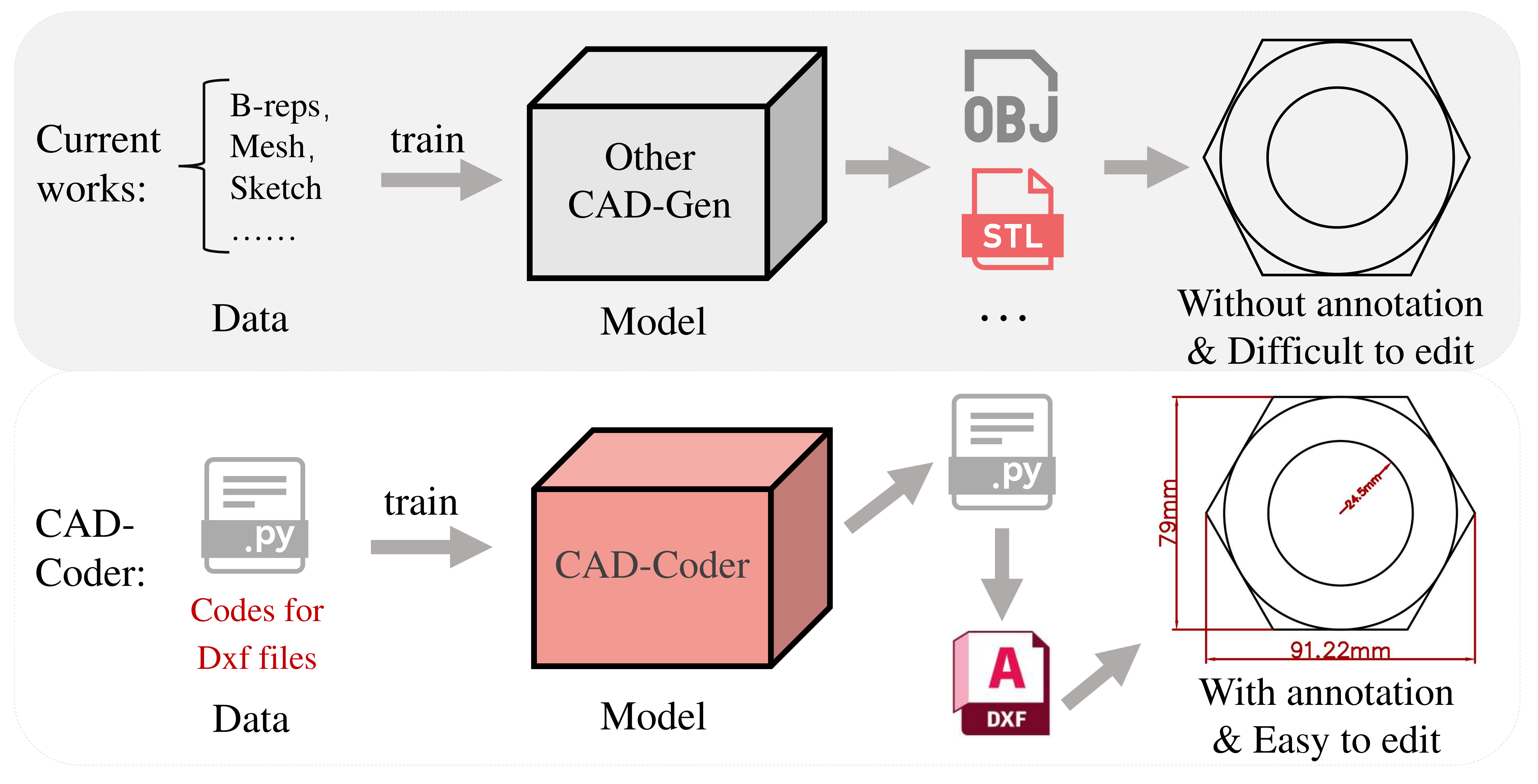}
   \caption{\textbf{Comparison of CAD-Coder with current works. }Comparing with other CAD generation methods, CAD-Coder uses a different form of dataset and produces more easily editable and annotated CAD models.}
   \label{s}
\end{figure}

Early research primarily focused on reconstructing CAD models from 3D point clouds\cite{structurenet,learning,pointflow,learningrepresentations}, or generating CAD models based on CAD command sequences\cite{sketchgen,deepcad,vq}. However, due to the high complexity of the above data format, these methods cannot guide the generation of CAD models through natural language, resulting in their poor practicability. Consequently, recent scholarly efforts have shifted towards text-guided CAD model generation \cite{llm4cad,alrashedy2024generating,Cad-llm,text2cad}, enabling users to obtain desired CAD models through natural language descriptions. Despite these advances, the outputs of these models fail to produce underlying universal CAD format files, making them still challenging for engineers to utilize in practical applications as shown in Figure~\ref{s}.

\begin{figure*}[htbp] 
  \centering
\includegraphics[width=\textwidth,height=\textheight,keepaspectratio]{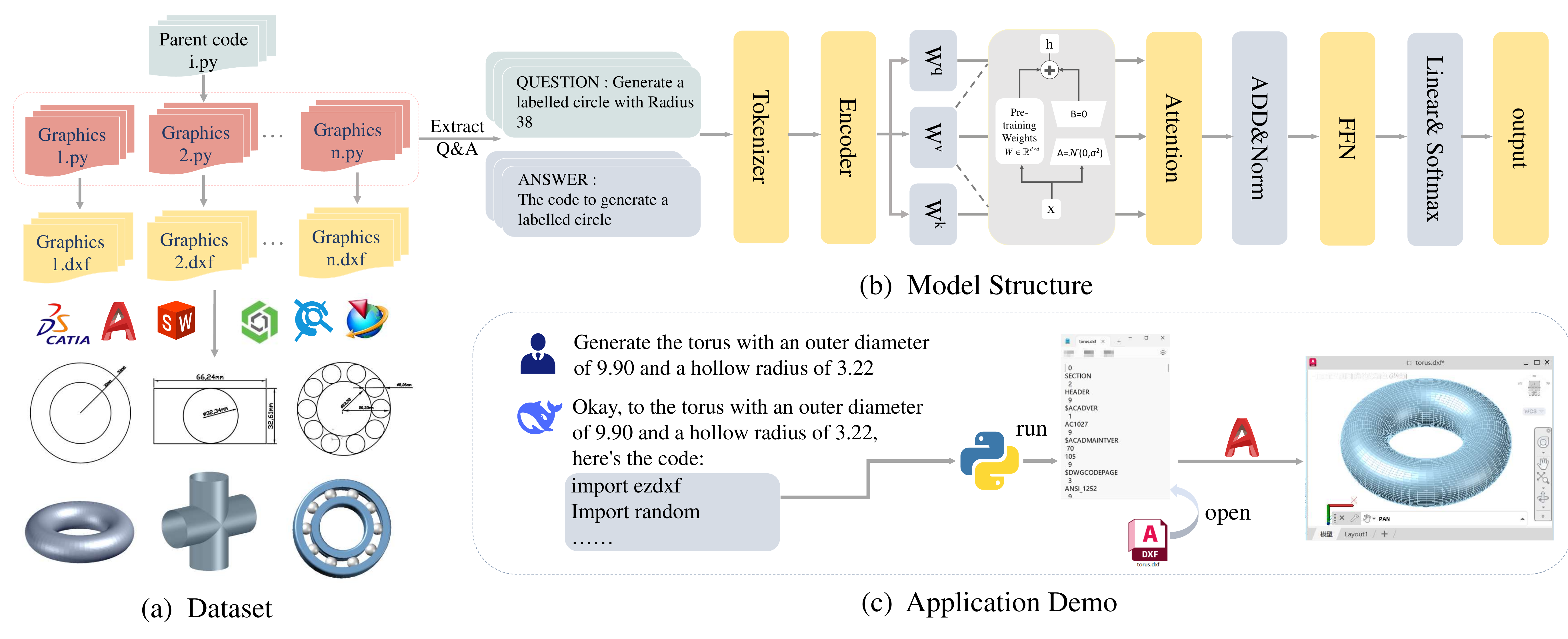}
  \caption{\textbf{Pipeline of the CAD-Coder.} By randomly assigning values to the free parameters in the parent code, a series of script codes along with their corresponding Dxf files are generated, forming the CFSC Dataset. The dataset contains both 3D models and 2D sketches, especially contains annotated data. The codes in dataset are matched with their corresponding natural language descriptions, from which relevant 
  question-answer pairs were extracted and injected into the DeepSeek-R1-Distill-Llama-8B model. Through training with the LoRA method, the model acquired the ability to infer CAD script code. Users can query the model to obtain the desired CAD script code, which can then be executed to generate Dxf files. These files can be opened and edited on various platforms.}
  \label{fig:short}
\end{figure*}

Inspired by large language models(LLMs) such as ChatGPT \cite{achiam2023gpt}, Llama \cite{llama3,touvron2023llama2openfoundation,llama} and code generation tasks like CodeLlama\cite{codellama} , Codex\cite{codex}, CodeGeeX\cite{CodeGeeX}, Phi\cite{abdin2024phi3technicalreporthighly}, etc., we propose CAD-Coder, a Python-based network for generating CAD sketches and models, as shown in Figure~\ref{fig:short}. By writing Python script  code with the ezdxf library, we construct easily editable underlying universal CAD format files (.Dxf) and achieve cross-modal generation from natural language to script code, and ultimately to CAD model. The resulting Dxf files of the CAD models
\cite{DXF} can be directly opened by common CAD platforms such as AutoCAD\cite{Autocad}, SolidWorks\cite{Solidworks}, CAXA\cite{CAXA}, CATIA\cite{Catia}, UG\cite{UG}, Onshape\cite{Onshape} and others.

Designing components and products requires precise geometric design and structural analysis. For practical CAD models, data annotation\cite{CAD-based} such as length, radius, angle, tolerance, chamfer and surface roughness are indispensable. However, current models are largely incapable of generating CAD models with annotation information, primarily due to limitations in existing datasets. To address this, we construct a  dataset, CAD Files Script Code (CFSC) Dataset, which leverages Python’s ezdxf library\cite{ezdxf} to construct CAD models. This dataset consists of Python script codes is capable of constructing CAD files alongside their corresponding CAD file outputs.
To sum up, the main contributions of this work include: 

\begin{itemize}

   \item We propose CAD-Coder, a CAD generation model capable of producing easily editable CAD models based on textual input. 

   \item Leveraging the annotation features of the ezdxf library for CAD sketches, our model enables accurate data annotation for 2D sketches. 

   \item We introduce the CFSC dataset, which includes a large number of CAD models along with their corresponding script codes and natural language descriptions.

\end{itemize}
\section{Related Work}
\label{sec:Related Works}

\textbf{CAD generative model.} Early work on sketch generation primarily focused on 2D sketch generation. Models such as SketchGen\cite{sketchgen} and CAD-as-language\cite{Computer-aided} utilized the Transformer\cite{transformer} architecture to handle geometric constraints in 2D sketches for generating 2D sketches. Later, models like DeepCAD\cite{deepcad}, SkexGen\cite{SkexGen}, and Draw Step by Step\cite{Draw} advanced the field by outputting 3D CAD operation sequences based on the Transformer architecture, marking a significant step toward generative CAD in the 3D domain. Additionally, the BrepGen\cite{brepgen} model directly generated 3D models in B-rep format through structured latent geometric representations. However, none of these models possess the capability to generate CAD sketches or models based on natural language. The latest advancements in CAD generation research, such as Text2CAD\cite{text2cad} and LLM4CAD\cite{llm4cad}, leveraged LLMs to achieve text-guided CAD generation. However, these approaches are constrained by dataset limitations, lacking the capability to produce diverse models. Additionally, they are unable to generate annotated sketches.

\noindent\textbf{CAD sketch and program synthesis.} A CAD sketch typically consists of one or more closed graphs (loops), with each loop composed of multiple primitive geometric elements such as lines, arcs, circles and so on. Designers can assist in sketch design through commands or programming languages\cite{parametric1,automatic}. Among existing methods, AutoLisp\cite{AutoLisp} and FreeCAD\cite{FreeCAD} commands are relatively popular. Autolisp\cite{AutoLisp}, as the built-in scripting language for AutoCAD\cite{Autocad}, offers strong interactivity but is limited in cross-platform applications. The FreeCAD\cite{FreeCAD} command interface is open-source and excels in geometric processing capabilities, yet its efficiency in handling complex Dxf files needs improvement\cite{DXF}. Both methods are constrained by platform limitations, lacking strong universality and requiring a high level of expertise.

\noindent\textbf{CAD dataset.} Existing CAD datasets can be categorized into two types: 2D sketches and 3D models. In the realm of 2D sketches, datasets such as SketchGraphs\cite{sketchgraphs}, Vitruvion\cite{vitruvion} and CAD-as-language\cite{Computer-aided} have established structured representation Fusion 360 Gallery\cite{fusion}, ABC\cite{abc}, Thingi10K\cite{thingi10k}, and CC3D\cite{pvdeconv,cadops} accomplish model generation tasks through CAD construction sequences or B-Rep formats. However, these datasets generally lack paired text descriptions and CAD models, leading to limitations in text-driven generation tasks. Although Text2CAD\cite{text2cad} pioneered the construction of a cross-modal dataset linking text prompts with CAD command sequences, existing datasets still share a common flaw: 2D sketches lack explicit geometric annotations.

\section{Methodology}
This section details the framework design methodology of the proposed CAD-Coder. We divide this section into three parts, including the core framework of CAD-Coder for CAD model generation (Sec.~\ref{CAD underlying universal file}), the construction process of our CFSC Dataset (Sec.~\ref{CFSC dataset}) and the elaboration on the training strategy of CAD-Coder in detail(Sec.~\ref{Model Architecture}). 

\subsection{CAD underlying universal file} \label{CAD underlying universal file}
As mentioned above, current CAD generation models are unable to produce editable CAD models. To address this issue, we use Dxf files as the final output format for CAD-Coder. Dxf file is a underlying universal format for Computer-aided design, primarily used to store and exchange 2D or 3D design data, known for their high precision and editability\cite{DXF}. However, since Dxf files, result in lengthy text formats that are complex and cumbersome after parsed, they are challenging to generate directly. Consequently, CAD-Coder leverages Python script codes to generate Dxf files, which is also adapted to the working principle of LLMs.

\subsection{CAD Files Script Code Dataset}\label{CFSC dataset}
\begin{figure}[h]
  \centering
    \includegraphics[width=\linewidth]{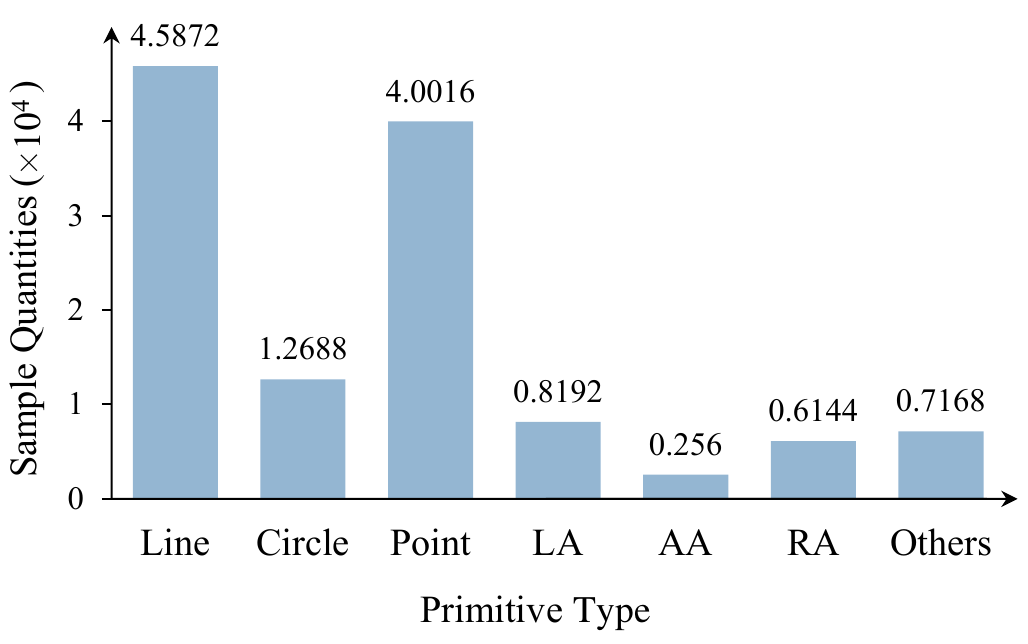}
   \caption{\textbf{Quantities of Different Primitive Types.} LA stands for linear annotation, AA is angle annotation, and RA is radius annotation.}
   \label{3}
\end{figure}

To better train CAD-Coder for generating annotated CAD models, we introduce the  CFSC Dataset, which comprises 29,130 Dxf files of CAD models along with their corresponding script codes. The dataset includes 2D sketches without annotations, 2D sketches with annotations, and 3D models. The statistical details of geometric primitives and data annotations in the dataset are presented in the Figure~\ref{3}.

\begin{figure}[b]
  \centering
    \includegraphics[width=\linewidth]{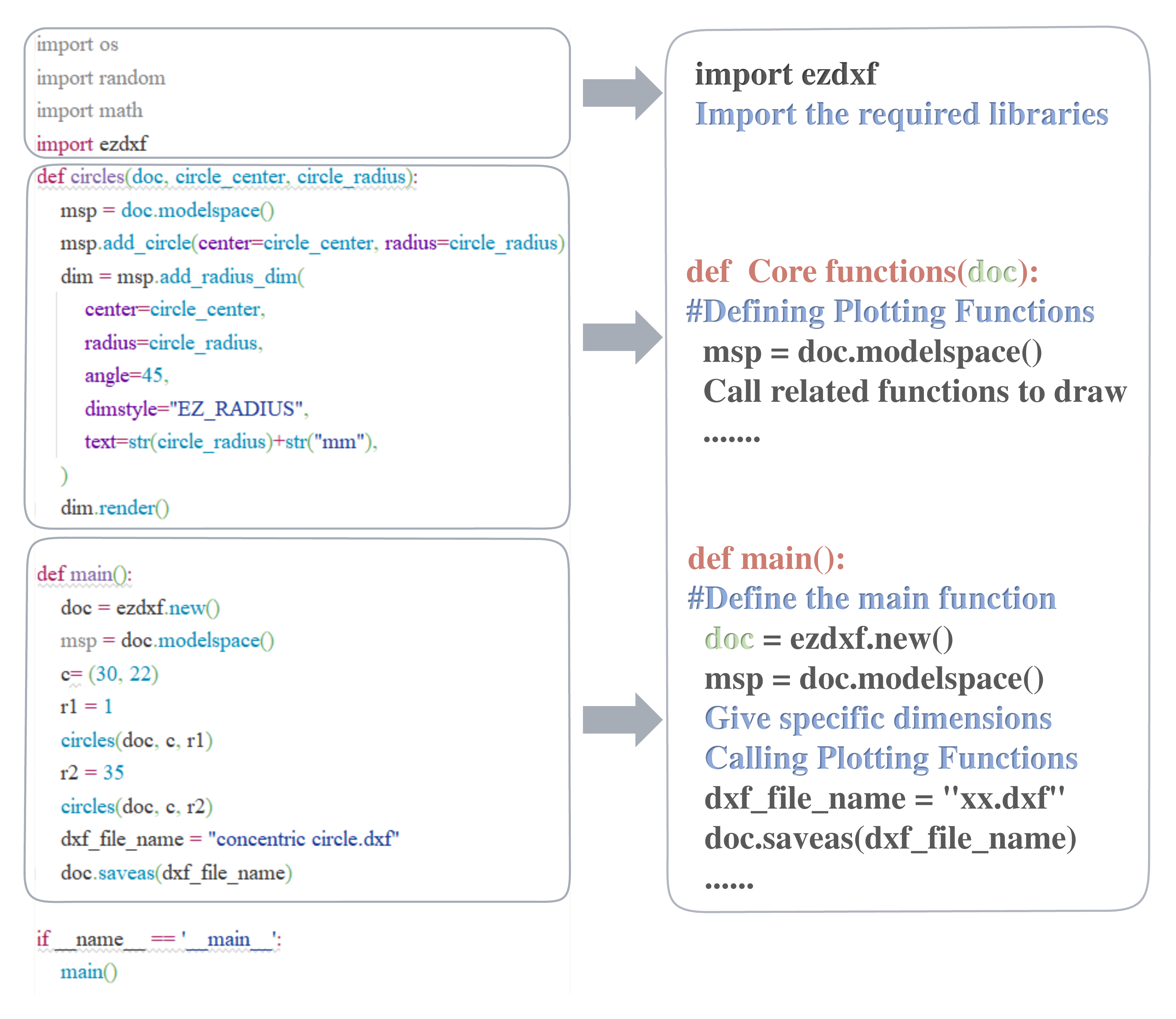}
   \caption{\textbf{Script code structure illustration.} The script code for each Dxf file consists of: (1) importing library functions, (2) the model construction function, and (3) the main function. The model construction function is responsible for defining the script code framework, including both model construction and adding annotations. The main function is used to call the model construction function, define the required dimensional parameters.}
   \label{4}
\end{figure}

Specifically, for a given shape, we first develop a framework script code \textbf{P} that incorporates all the necessary constraint relationships for the shape. However, instead of providing actual dimensional information, \textbf{P }references a set of parametric variables {$v$1, $v$2, ..., $v$n} to represent the fundamental characteristics of the shape, such as dimensions and position. Subsequently, we employ a randomization algorithm R to assign values to the parameter set {$v$1, $v$2, ..., $v$n}, thereby generating diverse shape script codes. It is important to note that the randomization algorithm R must account for the legality of the shape. For example, in the case of a hexagonal nut, the nominal diameter \textit{D} must be smaller than the short diameter \textit{De} of the hexagon. Clearly, such randomization algorithms are not uniform across different shapes. We will list the constraints required for various shapes in supplementary material.

By repeatedly sampling the parameter set, we can generate a series of subscript codes $p$1, $p$2, ..., $p$n, each corresponding to a different combination of parameters. These subscript codes independently produce their respective Dxf files. To better adapt to the code generation task, we standardize the structure of script codes, as shown in Figure~\ref{4}.
\begin{figure}[h]
  \centering
    \includegraphics[width=\linewidth]{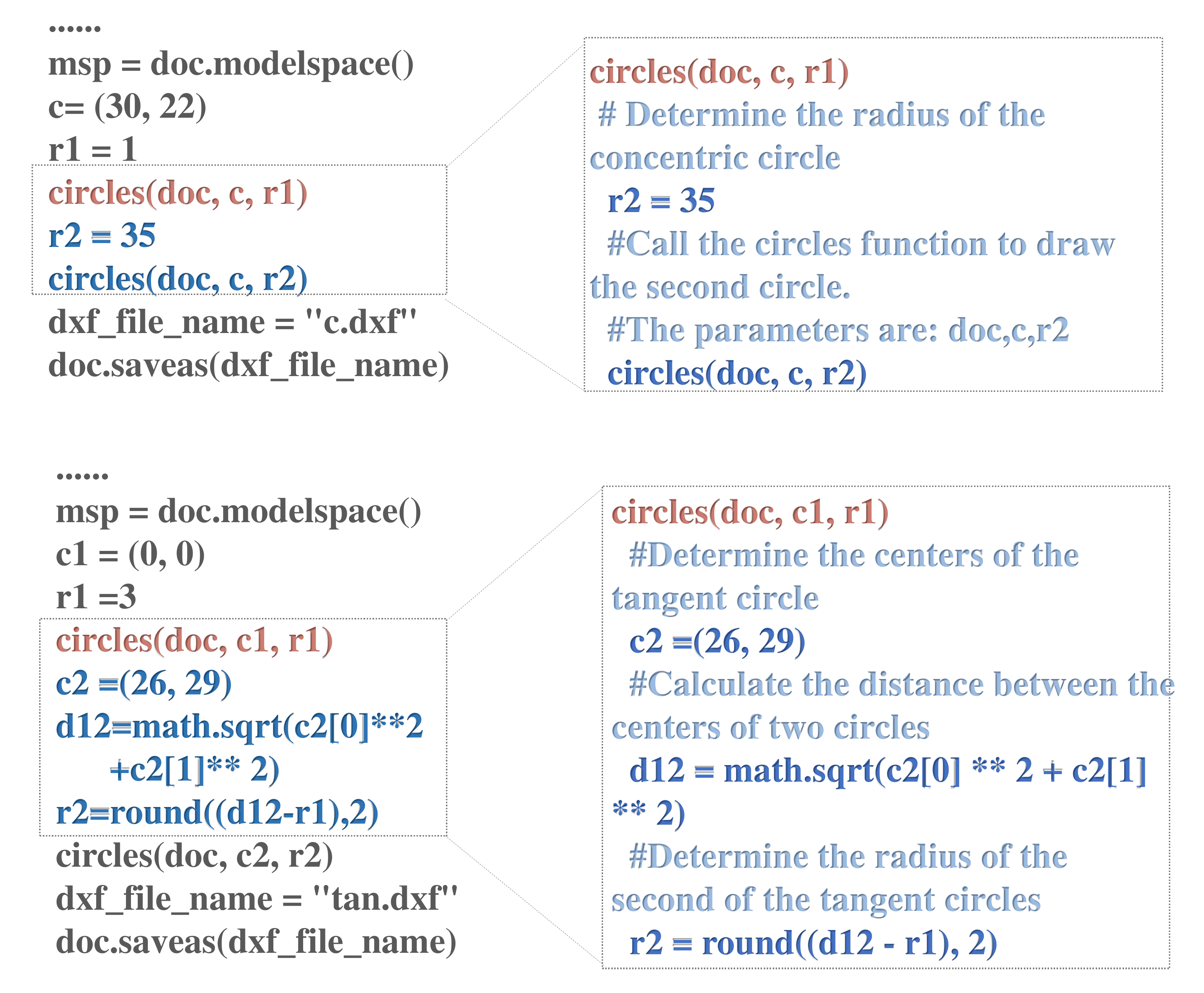}
   \caption{\textbf{Handling of similar script code segments.} Comments are added to script code segments that are prone to confusion, making it easier for the model to differentiate between these data.}
   \label{5}
\end{figure}
The unified script code structure results in high similarity in script codes for certain shapes. For example, the script codes for concentric circles and tangent circles differ only minimally in the main function, as shown in Figure~\ref{5}. Such minor differences are prevalent and difficult to capture during model training, leading to instability in the model's performance when generating script codes. When the model infers ``circles(doc, c, r1)'', it cannot determine whether the next line should generate $\text{r}2=35$ or $\text{c}2=(26,29)$. To address this issue, we use the LLM to add comments to the script codes, particularly in areas prone to such discrepancies. In fact, annotating the parent script code \textbf{P} alone suffices to annotate an entire set of data, significantly reducing the workload.

\subsection{Model Architecture}\label{Model Architecture}
\textbf{Data preprocessing.}
To enable the LLMs to adapt to the CAD generation task and learn the script codes patterns better in our dataset, we need to perform data preprocessing by extracting `Natural language description - Python script code' question-answer pairs from the dataset. Specifically, we create corresponding script codes for each CAD model and provide appropriate prompts, forming sets of question-answer pairs $\langle q_{i},a_{i}\rangle$, where $q_{i}$ is the natural language description of model $i$, and 
 $a_{i}$ 
 is the script code of model \textit{i}. Subsequently, we use the Byte Pair Encoding (BPE) algorithm to tokenize the data, converting the original questions and script code snippets into a series of token sequences:
\begin{equation}q_i=(t_{i,1},t_{i,2},...,t_{i,Lq},c_{i,1},c_{i,2},...,c_{i,La}),\end{equation}
where $L_{q}$ and $L_{a}$ represent the number of tokens for the question and answer respectively, $t$ and $c$ represent the tokens of the question and the answer, respectively.

Then, the tokens are converted into vectors through an embedding matrix \textit{E}:
\begin{equation}
\begin{split}
x_i=E(q_i)=(e(t_{i,1}),e(t_{i,2}),...,e(t_{i,Lq}),\\
e(c_{i,1}),e(c_{i,1}),...,e(c_{i,La})),
\end{split}
\end{equation}
 $x_{i}$ is an encoded vectors that We will inject into the DeepSeek-R1-Distill-Llama-8B model to guide the generation of CAD models. 

\noindent\textbf{Distillation model.}
We select the DeepSeek-R1-Distill-Llama-8B\cite{liu2024deepseek} model as the base model. Compared to the standard Llama model, the distilled model transfers the reasoning capabilities of DeepSeek-R1 to a more lightweight 8B parameter model through knowledge distillation techniques, resulting in enhanced reasoning abilities. The primary method of distillation involves minimizing the Kullback-Leibler Divergence (KL Divergence) by comparing the output distributions of the teacher model and the distilled model:
\begin{equation}\mathcal{L}_{\mathrm{distill}}=\sum_xD_{\mathrm{KL}}(P_{\mathrm{teacher}}(y|x)\parallel P_{\mathrm{student}}(y|x)),\end{equation}
where $P_{teacher}$ is DeepSeek-R1, and $P_{student}$ is Llama-8B, \textit{x} and \textit{y} refer to their respective inputs and outputs. 

Then we smooth the output of the teacher model by temperature scaling to avoid overfitting the hard labels of the teacher model. 
\begin{equation}P_{\mathrm{teacher}}^\tau(y|x)=\mathrm{softmax}\left(\frac{\log P_{\mathrm{teacher}}(y|x)}{\tau}\right),\end{equation}

The temperature parameter $\mathrm{\tau}$ softens the distribution, making it easier for the student model to learn the implicit reasoning logic of the teacher model.

After distillation, the model further optimizes the quality of generation by combining format rewards with accuracy rewards for continued reinforcement learning (RL). DeepSeek-R1-Distill-Llama-8B possesses inference performance close to that of the full version of the R1 model, better aligning user-input cue words with the model's output content.

\noindent\textbf{LoRA.} To avoid significant computational and storage overhead, we employed the LoRA method during the model's learning process. The LoRA method reduces the number of parameters that need to be updated through low-rank matrix decomposition. 
Specifically, When injecting LoRA into the self-attention layer of the Transformer, the LoRA parameters are:
\begin{equation}W^{\prime}=W_0+\Delta W=W_0+BA,\end{equation} 
where $W\in\mathbb{R}^{d\times k}$ is original weight, $\Delta W\in\mathbb{R}^{d\times k}$ is a low-rank matrix, $A\in\mathbb{R}^{r\times k}$, $B\in\mathbb{R}^{d\times r}$, $\mathrm{r}$ is the low-rank dimension, which is usually much smaller than $\mathrm{d}$ and $\mathrm{k}$. During the fine-tuning process, only the low-rank matrices $\mathrm{A}$ and $\mathrm{B}$ need to be updated, while the original weight matrices remain unchanged.

\section{Experiment}
In this section, we first propose four novel metrics for measurement (Sec.~\ref{Metrics}). Then we conduct qualitative and quantitative evaluations of the CAD-Coder from three aspects, including script codes, 2D sketches without annotations, 2D sketchs with annotations and 3D models (Sec.~\ref{Experiment on generative ability.}). Additionally, we perform comparative experiments between CAD-Coder and existing state-of-the-art LLMs to demonstrate the comprehensive capabilities of CAD-Coder (Sec.~\ref{Model Comparison and Analysis}). Furthermore, we highlight the model's exceptional performance in cross-platform compatibility (Sec.~\ref{Cross-Platform Capability}). More experiment details can be found in the supplementary material.

\subsection{Metrics}\label{Metrics}

To better evaluate the performance of different CAD models, we employ the following evaluation metrics in the experiments:

\noindent\textbf{Function accuracy (ACC-F). }Compare the generated script code with the ground truth, if the set of functions composed of all functions in the generated script code is equal to the set of functions composed of all functions in the ground truth, then it is defined as the same function of the generated script code, and the overall generative framework of the model can be judged by testing the rate of the same function of all script codes. 
\begin{equation}ACC-F=\frac{\sum_{j=1}^{N_c}\sum_{i=1}^{N_j}I\left[f_i=\hat{f}_i\right]}{\sum_{j=1}^{N_c}N_j},\end{equation}
$N_{c}$ denotes that there are $N_{c}$ sets of data, $N_{j}$ denotes the number of functions contained in each set of data, $f_{i}$ and $\hat{f}_{i}$ is the set of functions in the ground truth and generated script code, respectively. \textit{I }is the indicator function, which takes the value 1 when the function names are the same.

\noindent\textbf{Parameter accuracy (ACC-P). }Compare the parameter number of each correctly generated function in the generated script code with the ground truth to get the parameter correctness of the generated script code, thus judging the accuracy of the details of the generated sketches and models.
\begin{equation}ACC-P=\frac{\sum_{p=1}^{N_e}\sum_{i=1}^{N_e}I[p_i=\hat{p}_i]}{\sum_{p=1}^{N_e}N_p},\end{equation}
$N_{a}$ indicates that there is $N_{c}$ function, $N_{p}$ indicates the number of parameters contained in each function, $p_{i}$ and $\hat{p}_{i}$ are the parameters of the function in the ground truth and generated script code, respectively. 

\noindent\textbf{Graphic accuracy (ACC-G). }It is obviously not enough to judge the accuracy of the script code, we need to check whether the image is standard or not, we define ACC-G to evaluate whether the generated image is accurate or not.
\begin{equation}ACC-G=\frac{\sum_{i=1}^DI_{\text{the graph is comect}}(x_i)}{D},\end{equation}
where \textit{D} denotes the number of all Dxf file script codes that can be compiled, and \textit{I} is an indicator function that takes the value 1 when the generated graph xi is correct.

\noindent\textbf{Annotation accuracy (ACC-A). }In order to test the model's annotation ability, we define the annotation accuracy ACC-A.
\begin{equation}ACC-A=\frac{\sum_{i=1}^DI_{\text{the annotation is correct}}(x_i)}{D},\end{equation}
where the indicator function \textit{I} takes the value 1 when the generated graph $x_{i}$ is annotated correctly.

\subsection{Experiment on Generative Ability }\label{Experiment on generative ability.}
\begin{figure}[h]
  \centering
    \includegraphics[width=\linewidth]{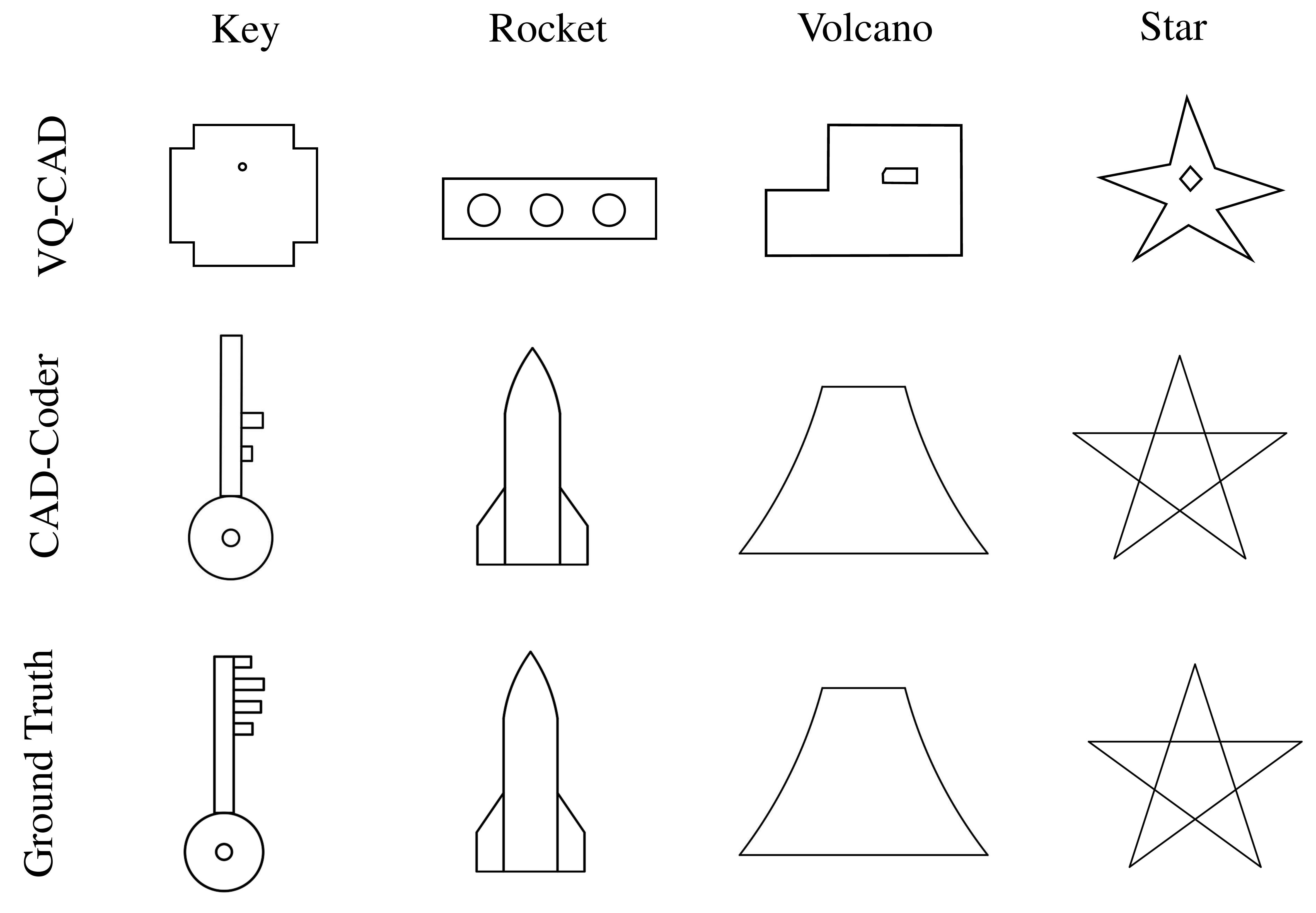}
   \caption{\textbf{Comparative evaluation of CAD-Coder and VQ-CAD.} It includes four kinds of contents: key, rocket, volcano and pentagon. In the figure, The first row is the ground truth, the second row is the result of VQ-CAD, and the third row is our result.}
   \label{6}
\end{figure}

\textbf{2D sketches without annotations.} Figure~\ref{6} demonstrates CAD-Coder's capability in generating abstract images without annotations, compared with VQ-CAD\cite{vq}. As shown in the Figure~\ref{6}, CAD-Coder can generally produce images that are roughly similar to the ground truth, and its generation performaance is superior to that of VQ-CAD.

\begin{figure}[t!]
  \centering
    \includegraphics[width=\linewidth]{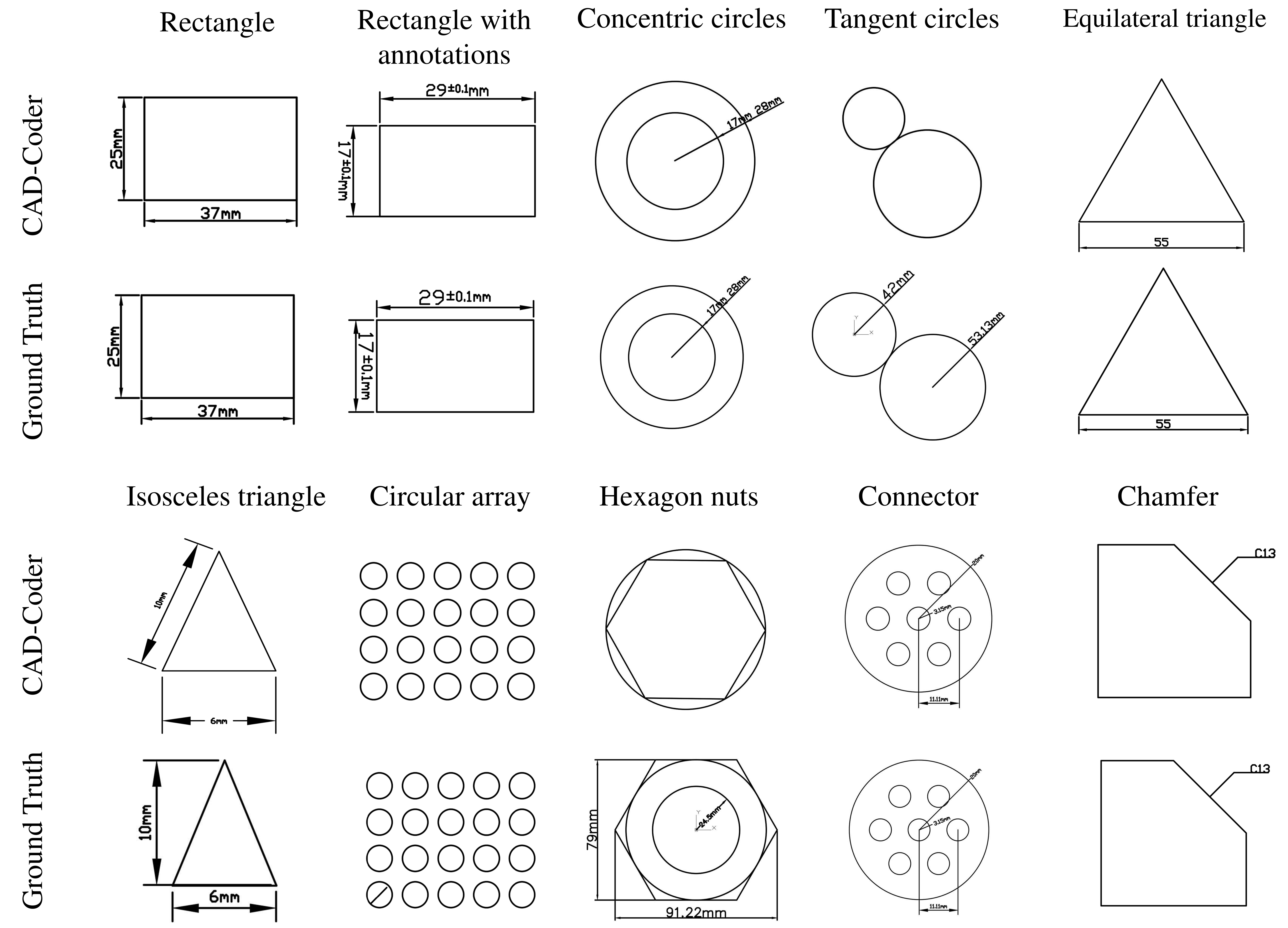}
   \caption{\textbf{Comparison of annotated sketches generated by CAD-Coder with ground truth.} There are 10 sets of comparative experimental results, and the results generated by CAD-Coder correspond to the ground truth.}
   \label{7}
\end{figure}

\noindent\textbf{2D sketches with annotations.}
Figure~\ref{7} demonstrates CAD-Coder's capability in generating annotated CAD models. While some minor details may not perfectly match the ground truth, the model largely reproduces the main content of the ground truth CAD sketches, and the annotations are relatively accurate.

\begin{figure}[h]
  \centering
    \includegraphics[width=\linewidth]{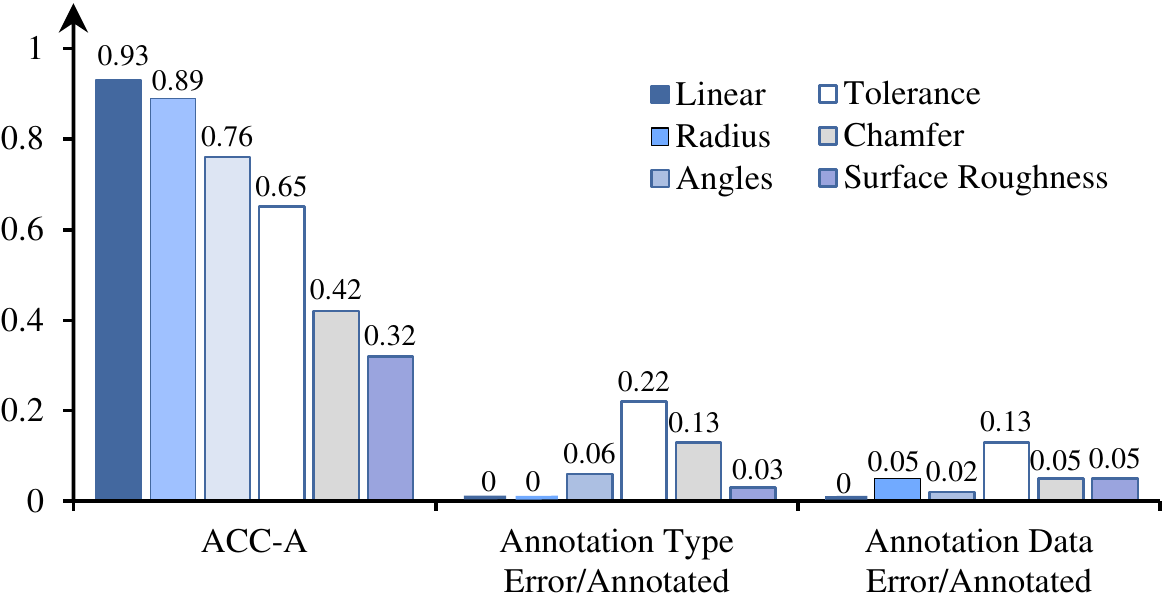}
   \caption{\textbf{Annotation capability assessment.} The bar chart includes three evaluation metrics: ACC-A (the probability of successful annotation), Annotation Type
Error/Annotated (the percentage of correct annotation types in the annotation) and Annotation Data Error/Annotated (the percentage of correct data in the annotation) were tested on six annotation types, including linear annotation, radius annotation, angle annotation, tolerance annotation, chamfer annotation and surface roughness annotation.}
   \label{8}
\end{figure}

Figure~\ref{8} demonstrates the variety of annotation types our model can handle, including tolerances, surface finishes, chamfers, angles, and more. We also evaluate the accuracy of the model's annotations. As shown in Figure~\ref{7}, for simple annotation types such as linear annotations and radius annotations, our model performs well and rarely makes errors in annotation types. However, for angle annotations, a small number of results mistakenly annotate diameters as radius, leading to a higher annotation data error.

For more complex annotations, such as chamfers, tolerances, and surface roughness, the CAD-Coder's sensitivity is slightly lower, but it still exhibits some capability. Tolerance annotations are prone to omissions or only annotating linear dimensions, resulting in a higher annotation type error. On the other hand, chamfers and surface roughness annotations, once applied, are generally accurate in type, demonstrating a certain level of stability.

\begin{figure}[t]
  \centering
    \includegraphics[width=\linewidth]{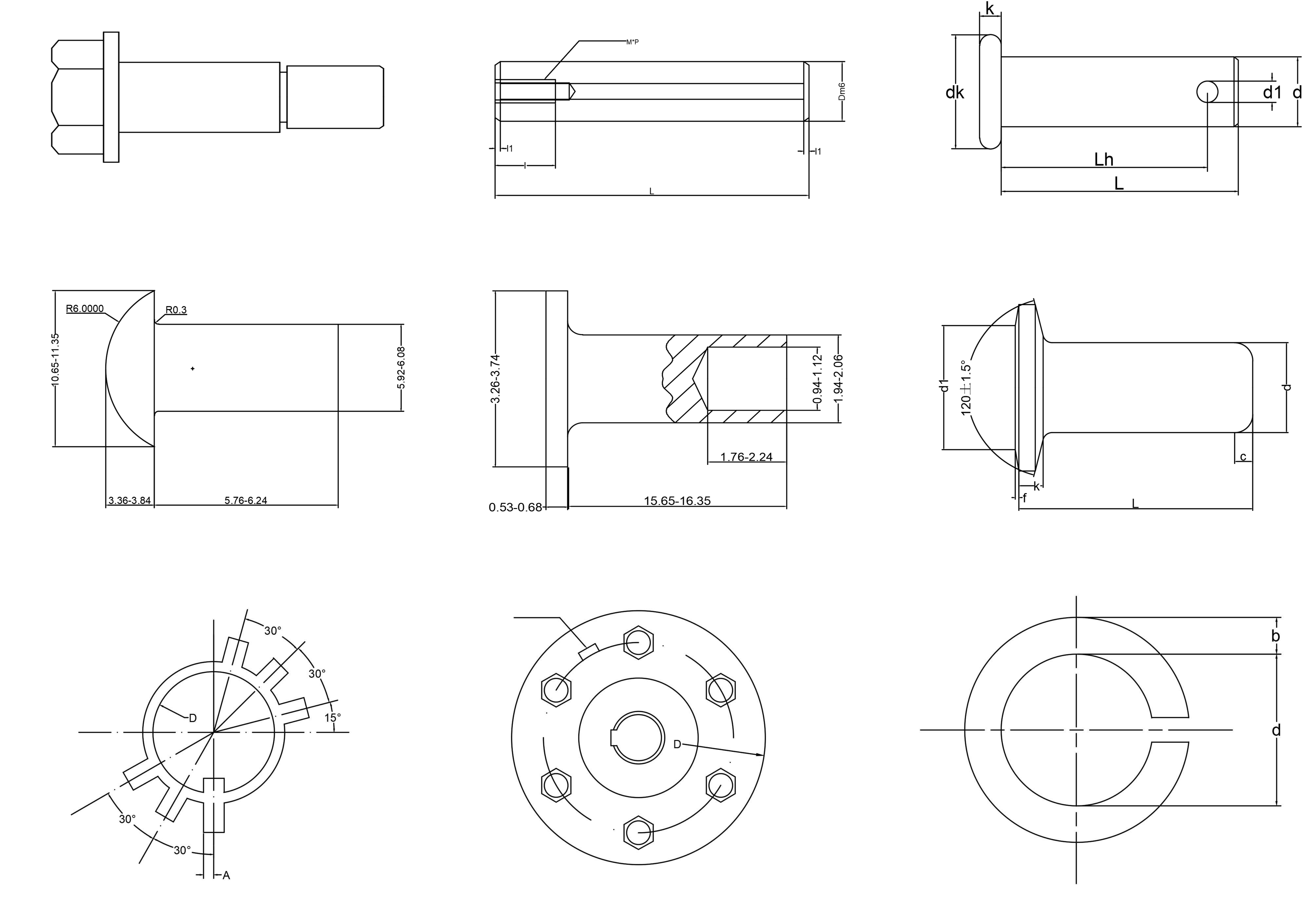}
   \caption{\textbf{More complex sketch generation.} The actual engineering parts containing various primitives and annotations are shown in this figure}
   \label{9}
\end{figure}

During the experiments, we observed that as the dataset and model scale increase, our method is capable of generating more complex CAD sketches, as illustrated in Figure~\ref{9}.

\begin{figure}[h]
  \centering
    \includegraphics[width=\linewidth]{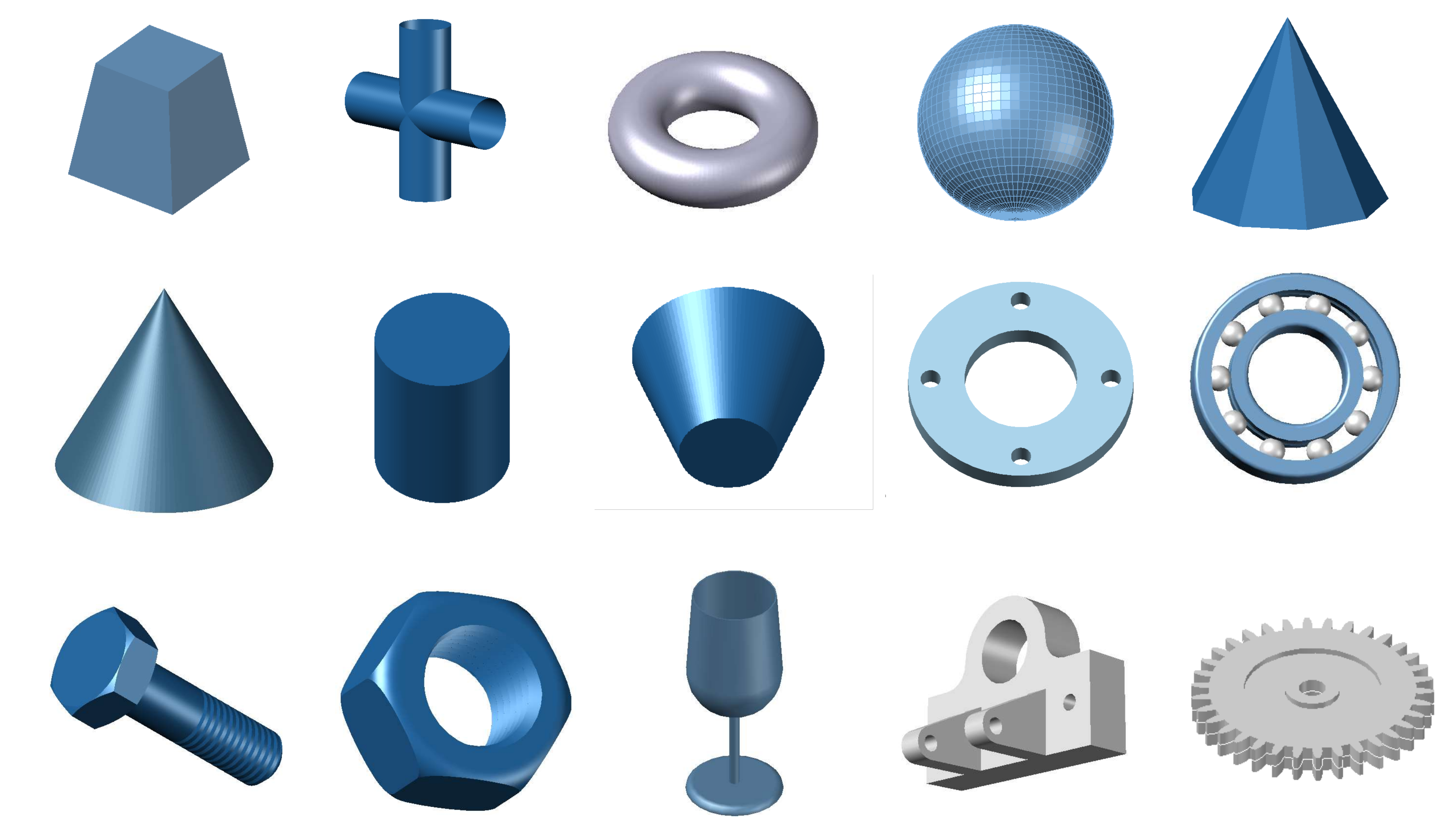}
   \caption{\textbf{Generated 3D models illustration.} CAD-Coder has a strong ability to generate 3D engineering parts including flanges, bearings, screws, nuts, gears, etc.}
   \label{10}
\end{figure}

\noindent\textbf{3D Models.} 
Although the Dxf format is primarily used for the design of 2D sketches, it also possesses strong capabilities for representing 3D models. We did not overlook this feature of Dxf files and created a substantial number of 3D models that can be opened with Dxf files, along with their corresponding code. After training, CAD-Coder was endowed with the ability to express 3D representations, as shown in Figure~\ref{10}.

\subsection{Model Comparison and Analysis}\label{Model Comparison and Analysis}
\begin{table*}[h]
\centering
\small

\setlength{\tabcolsep}{1.5mm} 
\begin{tabularx}{\textwidth}{l>{\centering\arraybackslash}X>{\centering\arraybackslash}X> {\centering\arraybackslash}X>{\centering\arraybackslash}X>{\centering\arraybackslash}X>{\centering\arraybackslash}X} 
\toprule
Model & pass@1$\uparrow$ & pass@3$\uparrow$ & pass@5$\uparrow$ & APR$\uparrow$  & ACC-G$\uparrow$ & ACC-A $\uparrow$\\
\midrule
Qwen2.5-Coder-14b\cite{yang2024qwen2} & 0.14   & 0.26   & 0.33   & 0.39   & 0.16  & 0.09  \\
ChatGPT-4\cite{achiam2023gpt}         & 0.16   & 0.33   & 0.39   & 0.49   & 0.17  & 0.03  \\
Deepseek-V3\cite{liu2024deepseek}       & 0.13   & 0.17   & 0.18   & 0.54   & 0.25  & 0.29  \\
Llama3.3-70b\cite{llama3}              & 0.17   & 0.23   & 0.24   &\textbf{0.82}    & 0.43  & 0.14  \\
CAD-Coder                      &\textbf{0.40}    & \textbf{0.74}   &\textbf{0.81}    & 0.79   &\textbf{0.68}   &\textbf{0.77 }   \\
\bottomrule
\end{tabularx}
 \caption{\textbf{Quantitative comparison with existing LLMs.} The evaluation metrics include Pass@k\cite{2021arXiv210703374C}, Average Parsing Rate (APR), Graphic Accuracy (ACC-G), and Annotation Accuracy (ACC-A). The best-performing model is highlighted in bold.}
 \label{table_2}
\end{table*}

\begin{figure}[htb]
  \centering

    \includegraphics[width=\linewidth]{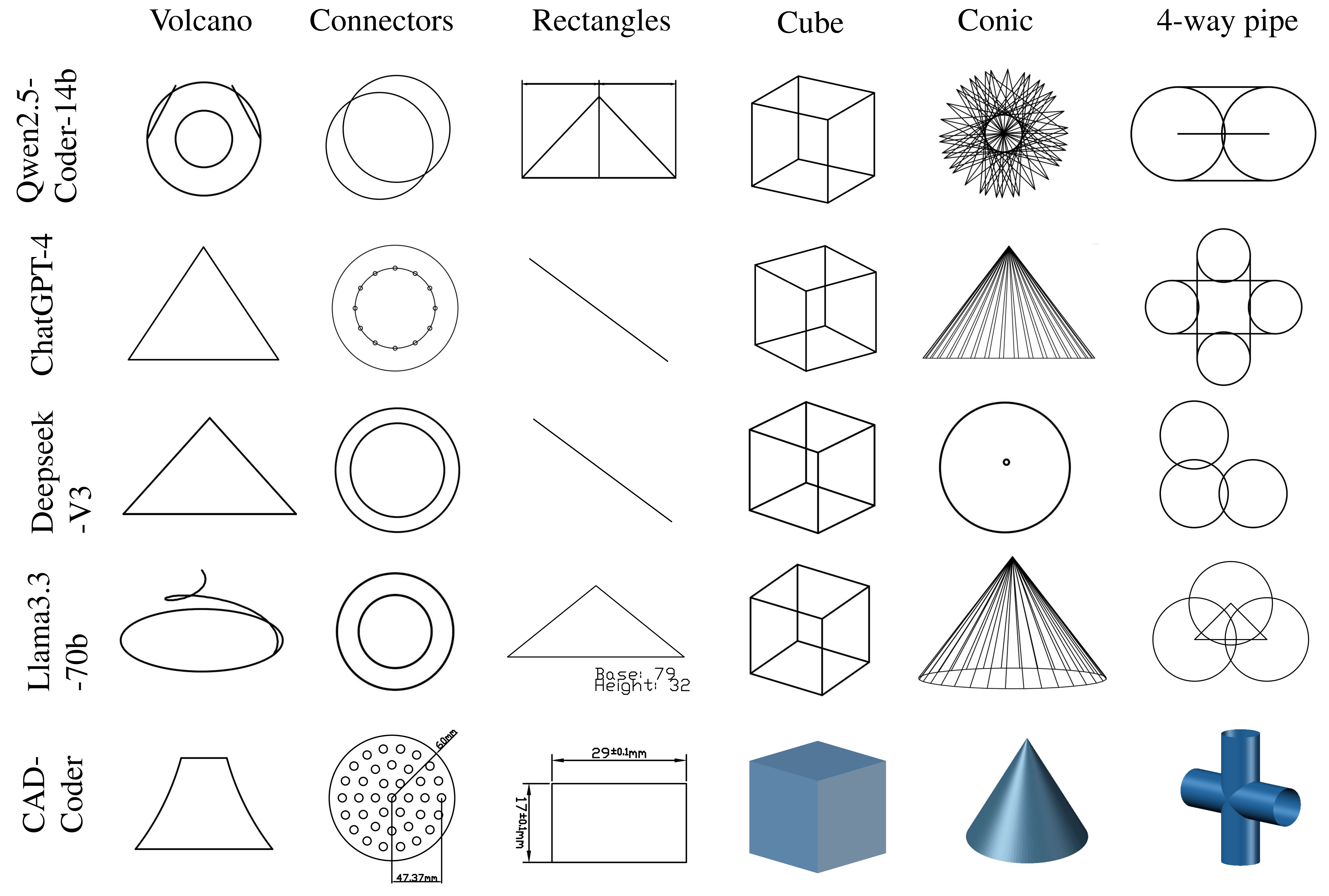}
    
   \caption{\textbf{Comparison of CAD-Coder's generation results with those of other LLMs.} A total of six sets of prompts are fed into the model: one set of unannotated 2D sketch prompt, two sets of annotated 2D sketch prompts, and three sets of 3D model prompts.}
   \label{11}
\end{figure}
To demonstrate the capability of our method in CAD script code generation tasks, we conduct extensive generation experiments and compared our results with existing state-of-the-art models. 
The experiments randomly selected 485 prompts, including 212 prompts for 3D models, 115 prompts for 2D sketches without annotations, and 158 prompts for 2D sketches with annotations. The results are shown in Table~\textbf{\ref{table_2}. }

\begin{table*}[htp]
\centering
\small
\setlength{\tabcolsep}{1.5mm} 
\begin{tabularx}{\textwidth}{l>{\centering\arraybackslash}X>{\centering\arraybackslash}X>{\centering\arraybackslash}X>{\centering\arraybackslash}X>{\centering\arraybackslash}X>{\centering\arraybackslash}X>{\centering\arraybackslash}X} 
\toprule
Method& ACC-F$\uparrow$ & ACC-P$\uparrow$& ACC-G$\uparrow$ & pass@1$\uparrow$ & pass@3$\uparrow$ & pass@5$\uparrow$&CD$\downarrow$   \\
\midrule
 CAD-Coder w/o. annotation    & \textbf{0.79}   &\textbf{0.83}   & \textbf{0.69}   & \textbf{0.42}   & \textbf{0.79}  & \textbf{0.89}  & \textbf{0.74}\\
 CAD-Coder w. annotation          & 0.66   & 0.76   & 0.51   & 0.33   & 0.59  & 0.74  & 0.88\\
\bottomrule
\end{tabularx}
\caption{\textbf{Comparison of annotated and non-annotated CAD-Coder generation ability. }The evaluation metrics include Pass@k\cite{2021arXiv210703374C}, Function Accuracy (ACC-F), Parameter Accuracy (ACC-P), Graphic Accuracy (ACC-G) and Chamfer Distance (CD).}
 \vspace{-5mm}
 \label{ttable2}
\end{table*}

Our method significantly outperforms other models across multiple metrics. The advantage in ACC-G demonstrates that our approach effectively translates natural language into geometric shapes, while the substantial lead in ACC-L proves that training with script code data for annotation generation is highly effective.

The Pass@k results indicate that the overall performance of our generated script codes are superior to that of other models. In terms of APR, the results of Llama are slightly better than our results, primarily because our method often generates complex but standard script codes, which have higher probability of failing to compile or run. In contrast, Llama tends to generate script codes that could compile and run but may not meet the required standards.

Figure~\ref{11} compares the CAD sketches generated by several models. The comparison clearly shows that our model generates CAD sketches with more accurate shapes, better understanding of prompts, and far superior annotation capabilities compared to other LLMs. Additionally, our model can generate more realistic 3D entities.

\subsection{Experiment on Cross-Platform Capability}\label{Cross-Platform Capability}

Existing program-driven CAD generation models typically use custom CAD commands as their output, which cannot be directly opened by common CAD platforms (e.g., AutoCAD\cite{Autocad}, SolidWorks\cite{Solidworks}, etc.). These models still require specific tools or scripts to process the command sequences to produce visual CAD results, resulting in limited cross-platform compatibility. 

\begin{figure*}[h]
  \centering
    \includegraphics[width=\linewidth]{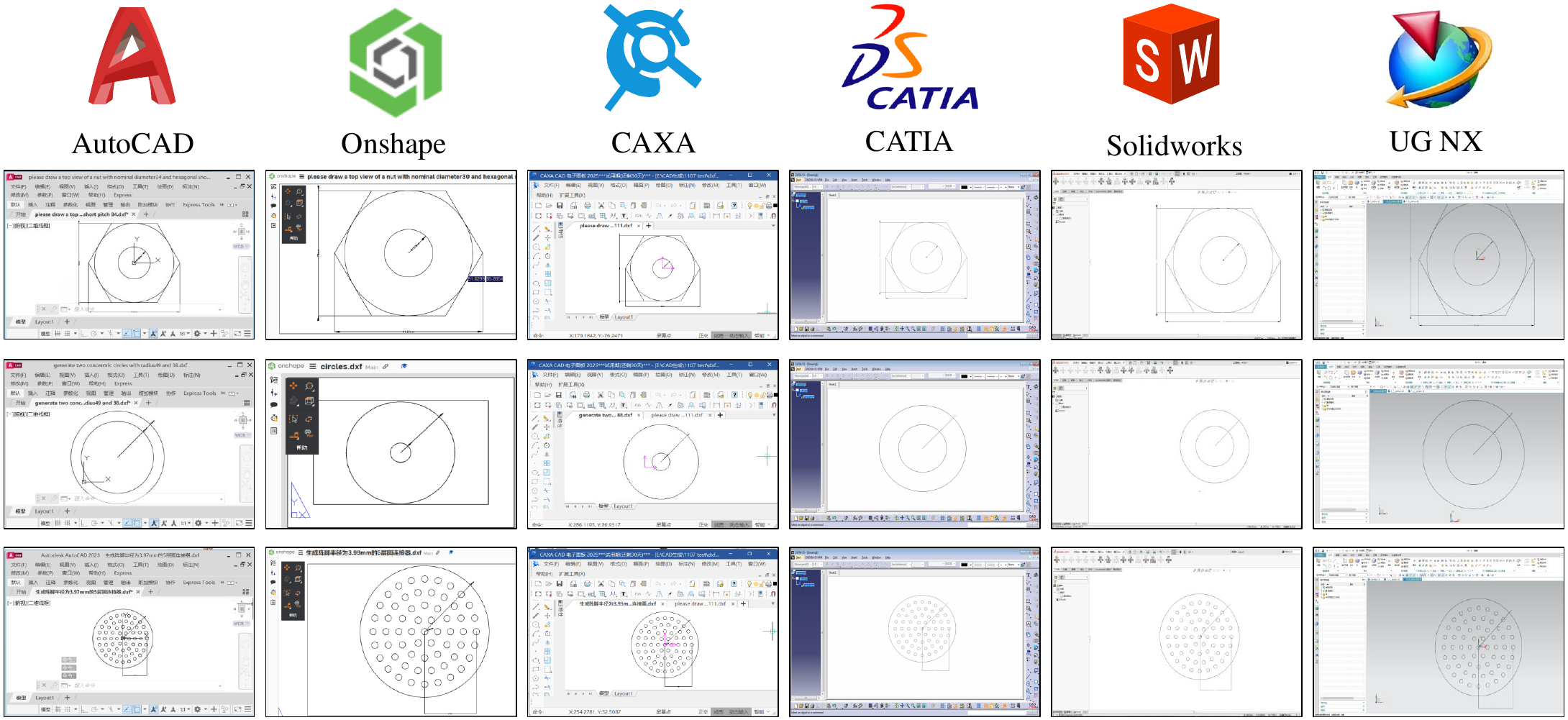}
   \caption{\textbf{Cross-platform capability illustration.} This figure displays the generated Dxf file being opened in different platforms/software, arranged from left to right as follows: AutoCAD\cite{Autocad}, Onshape\cite{Onshape}, CAXA\cite{CAXA}, CATIA\cite{Catia}, SolidWorks\cite{Solidworks}, and UG\cite{UG}.}
   \label{12}
\end{figure*}

Our model, however, starts with Python script codes. By running the generated Python script codes, it produces a universal underlying CAD file in Dxf format. The resulting Dxf files exhibit excellent cross-platform compatibility and can be opened in almost all mainstream CAD software and platforms, yielding the expected sketches and models, as shown in Figure\ref{12}.
\begin{table*}[htb]
\centering
\small

\setlength{\tabcolsep}{1.5mm} 
\begin{tabularx}{\textwidth}{l>{\centering\arraybackslash}X>
{\centering\arraybackslash}X>
{\centering\arraybackslash}X>
{\centering\arraybackslash}X> {\centering\arraybackslash}X>{\centering\arraybackslash}X>{\centering\arraybackslash}X>{\centering\arraybackslash}X} 
\toprule
Model & pass@1$\uparrow$ & pass@3$\uparrow$ & pass@5$\uparrow$ & APR$\uparrow$  & ACC-G$\uparrow$ & ACC-P $\uparrow$& ACC-F $\uparrow$& CD $\downarrow$\\
\midrule
CAD-Coder w. LoRA& \textbf{0.33}  & \textbf{0.59}   & \textbf{0.74}   & \textbf{0.79}  & \textbf{0.51} & \textbf{0.76}  & \textbf{0.66}  & \textbf{0.88}  \\
CAD-Coder w/o. LoRA& 0.25   & 0.47   & 0.59   & 0.75   & 0.47  & 0.70  & 0.52  & 0.93\\
\bottomrule
\end{tabularx}
 \caption{\textbf{Comparison of full parameter and LoRA fine-tuning.} The evaluation metrics include Pass@k\cite{2021arXiv210703374C}, Average Parsing Rate (APR), Graphic Accuracy (ACC-G), Function Accuracy (ACC-F), Parameter Accuracy (ACC-P), and Chamfer Distance (CD) The best-performing model is highlighted in bold.}
 \label{ttable3}
 
\end{table*}

\subsection{Ablation Study}
\label{sec:Ablation_Study}

In this section, we conduct ablation studies to demonstrate the effectiveness of the proposed CAD-Coder. More ablation experiments are given in the supplementary materials.

\noindent\textbf{Different Annotation Strategies.} To validate our model's unique capability in generating annotated CAD sketches, we conduct ablation studies comparing performance with and without annotation generation. By leveraging the ezdxf library's annotation features, CAD-Coder pioneers automated geometric annotation in 2D CAD generation. From Table\textbf{~\ref{ttable2}}, we can get that the generation without annotation is better than the generation with annotated content, both for the evaluation of the script codes alone and for the evaluation of the specific generated graphs. The performance gap mainly stems from the increased complexity in annotated scripts, which require handling additional geometric relationships and parameters. This highlights both the challenge and importance of accurate annotation generation, pointing to promising directions for future improvements.

\noindent\textbf{Effect of Fine-tuning Strategies.} Then, we conducted full parameter fine-tuning and LoRA fine-tuning on the distilled model to compare the effects of different methods on the model's generation capabilities. From Table\textbf{~\ref{ttable3}}, we can get that CAD-Coder with LoRA exhibit better performance in both the accuracy of script code generation and the accuracy of CAD model generation. The results indicate that the LoRA method is more suitable for our task, as our dataset is not very large, and the LoRA method helps reduce the occurrence of overfitting.The LoRA method consumes less time, storage, and resources to complete CAD model generation tasks, making it more convenient to train a user-friendly interactive CAD generation model.

\section{\textbf{Conclusion}}
In this paper, we introduced CAD-Coder, the first interactive model capable of generating annotated CAD files from natural language descriptions. We constructed the CFSC dataset, containing 29,130 Dxf files with corresponding script codes. \textbf{For anonymous reason, this dataset will be released upon acceptance of the paper.} In the future, we plan to expand the CFSC to encompass a more diverse range of engineering components and annotation types.

{
    \small
    \bibliographystyle{ieeenat_fullname}
    \bibliography{main}
}

\clearpage

\renewcommand\thesection{\Alph{section}}
\renewcommand\thefigure{\Alph{figure}}
\renewcommand\thetable{\Alph{table}}
\renewcommand\thesubsection{\thesection.\arabic{subsection}}

\newcommand*{\affaddr}[1]{#1} 
\newcommand*{\affmark}[1][*]{\textsuperscript{#1}}
\newcommand*{\email}[1]{\texttt{#1}}

\newcommand{\tabincell}[2]{\begin{tabular}{@{}#1@{}}#2\end{tabular}}

\twocolumn[
\begin{center}
	{\LARGE \textbf{Supplemental Materials\\~\\~\\}}
\end{center}]
\setcounter{section}{0}
\setcounter{table}{0}
\setcounter{figure}{0}

\hspace{-5mm}The content of this supplementary material involves:
\vspace{2mm}

	\hspace{-4mm}A. Experimental Setup and Costs in Sec.~\ref{sec:Experimental Setup and Costs}.

    \hspace{-4mm}B. Details of LoRA Fine-Tuning in Sec.~\ref{sec:Details of LoRA Fine-Tuning}.

    \hspace{-4mm}C. Evaluation Metrics in Sec.~\ref{sec:Evaluation Metrics}.

    \hspace{-4mm}D. More Ablation Studies in Sec.~\ref{sec:More Ablation Studies}.

    \hspace{-4mm}E. Details of Size Constraints in Sec.~\ref{sec:Details of Size Constraints}.

    \hspace{-4mm}F. Parent Code Example in Sec.~\ref{sec:Parent Code Example}.

\section{Experimental Setup and Costs}\label{sec:Experimental Setup and Costs}
We performed multi-round fine-tuning of the DeepSeek-R1-Distill-Llama-8B model on a dataset containing 29130 samples, approximately 2.27 million tokens in size, using a single NVIDIA V100 GPU. The learning rate was set to 0.0002, the batch size was set to 4, the sequence length was set to 1048, and the training was conducted for 2 epochs.
\section{Details of LoRA Fine-Tuning}\label{sec:Details of LoRA Fine-Tuning}
Specifically, for the L-th layer transformer, the LoRA increment for Query/Value is:
\begin{equation}Q=H_{l-1}(W_q+B_qA_q)^T=H_{l-1}W_q^T+H_{l-1}A_q^TB_q^T\end{equation}
Similarly, the value projection is:
\begin{equation}V=H_{l-1}(W_v+B_vA_v)^T\end{equation}
where $B_q^{(l)},B_v^{(l)}\in\mathbb{R}^{d\times r}$,$A_q^{(l)},A_v^{(l)}\in\mathbb{R}^{r\times d}$,$r\ll d$is the LoRA rank,$\Delta W\in\mathbb{R}^{d\times d}$ is the weight increment.
Attention is calculated as
\begin{equation}\mathrm{Attention}(Q,K,V)=\mathrm{softmax}\left(\frac{QK^\top}{\sqrt{d}}\right)V\end{equation}
Feedforward network (FFN): activated using SwiGLU, keeping the original parameters unchanged
\begin{equation}\mathrm{FFN}(X)=\mathrm{GeLU}(XW_1^{(l)})W_2^{(l)}\end{equation}
Residuals and Normalization:
\begin{equation}X_{\mathrm{out}}=\mathrm{LayerNorm}\left(X+\mathrm{Attention}+\mathrm{FFN}(X)\right)\end{equation}
\section{Evaluation Metrics}\label{sec:Evaluation Metrics}
\textbf{Pass@k. }Common evaluation metrics for code generation models refer to the criteria used to assess performance in a given generation task. If at least one of the k candidate results generated by the model meets the predefined success criteria (such as passing tests or satisfying specific conditions), the task is considered "successful." The calculation formula is as follows:
\begin{equation}pass@k:=E_{problems}\left[1-\frac{
\begin{pmatrix}
n-c \\
k
\end{pmatrix}}{
\begin{pmatrix}
n \\
k
\end{pmatrix}}\right]\end{equation}
The model generates \textit{n} (where $n > k$) pieces of code for each question, and then randomly selects \textit{k} pieces of code from these. If at least one of the \textit{ k} pieces of code passes the unit tests, it is considered successful. Here, \textit{c} represents the total number of pieces of code that can pass the unit tests.

We tested the results of pass@k using python code. 

In addition, we used the Comparecloud platform to convert Dxf files into point cloud format files, and tested the chamfer distance (CD) of the generated results with the help of python's open3d library.
\section{More Ablation Studies}\label{sec:More Ablation Studies}
We conducted three sets of ablation experiments to assess the impact of different datasets, methods, and baseline models on the generation performance of CAD-Coder.
\begin{table*}[htb]
\centering
\small

\setlength{\tabcolsep}{1.5mm} 
\begin{tabularx}{\textwidth}{l>{\centering\arraybackslash}X>
{\centering\arraybackslash}X>
{\centering\arraybackslash}X>
{\centering\arraybackslash}X> {\centering\arraybackslash}X>{\centering\arraybackslash}X>{\centering\arraybackslash}X>{\centering\arraybackslash}X} 
\toprule
Model & pass@1$\uparrow$ & pass@3$\uparrow$ & pass@5$\uparrow$ & APR$\uparrow$  & ACC-G$\uparrow$ & ACC-P $\uparrow$& ACC-F $\uparrow$& CD $\downarrow$\\
\midrule
CAD-Coder with Comments& 0.40   & 0.74   & 0.81  & 0.79   & \textbf{0.68}  & \textbf{0.76}  & 0.66  & \textbf{0.88}  \\
CAD-Coder without Comments& 0.21   & 0.54   & 0.62   & 0.49   & 0.47  & 0.54  & 0.43  & 1.31  \\
CAD-Coder with 3D models& 0.37   & 0.72   & 0.78  & 0.83   & 0.64  & 0.62  & 0.61  & 0.72  \\
CAD-Coder without 3D models&\textbf{ 0.45}   & \textbf{0.81}   & \textbf{0.90}   &\textbf{0.86}   & 0.43  & 0.75  & \textbf{0.68} & 0.84\\
\bottomrule
\end{tabularx}
 \caption{\textbf{Comparison of full parameter and LoRA fine-tuning.} The evaluation metrics include Pass@k[15], Average Parsing Rate
(APR), Graphic Accuracy (ACC-G), Function Accuracy (ACC-F), Parameter Accuracy (ACC-P), and Chamfer Distance (CD) The best-
performing model is highlighted in bold.}
 \label{table_5}
\end{table*}

\begin{table*}[htb]
\centering
\small

\setlength{\tabcolsep}{1.5mm} 
\begin{tabularx}{\textwidth}{l>{\centering\arraybackslash}X>
{\centering\arraybackslash}X>
{\centering\arraybackslash}X>
{\centering\arraybackslash}X> {\centering\arraybackslash}X>{\centering\arraybackslash}X>{\centering\arraybackslash}X>{\centering\arraybackslash}X} 
\toprule
CAD-Coder based on & pass@1$\uparrow$ & pass@3$\uparrow$ & pass@5$\uparrow$ & APR$\uparrow$  & ACC-G$\uparrow$ & ACC-P $\uparrow$& ACC-F $\uparrow$& CD $\downarrow$\\
\midrule
 Llama                       & 0.32   & 0.61   & 0.73  & 0.60   & 0.51  & 0.58  & 0.47  & 1.20  \\
 CodeLlama                   & 0.27   & 0.54   & 0,65  & 0.51   & 0.46  & 0.47  & 0.34  & 1.76  \\
 Qwen                        & 0.35   & 0.56   & 0.76  & 0.56   & 0.47  & 0.55  & 0.43  & 1.35  \\
 Qwen-Coder                  & 0.22   & 0.47   & 0.59  & 0.45   & 0.44  & 0.41  & 0.27  & 2.10 \\
 DeepSeek-R1-Distill-Llama-8B& \textbf{0.40}   &\textbf{ 0.74}   & \textbf{0.81}  & \textbf{0.79}   & \textbf{0.68}  & \textbf{0.76}  & \textbf{0.66}  & \textbf{0.88 } \\
 DeepSeek-R1-Distill-Qwen-7B & 0.38   & 0.64   & 0.76  & 0.68   & 0.59  & 0.66  & 0.54  & 0.95  \\
\bottomrule
\end{tabularx}
 \caption{\textbf{Quantitative comparison with existing LLMs.} The evaluation metrics include Pass@k\cite{2021arXiv210703374C}, Average Parsing Rate (APR), Graphic Accuracy (ACC-G), and Annotation Accuracy (ACC-A). The best-performing model is highlighted in bold.}
 \label{table_6}
\end{table*}

First, we trained the DeepSeek-Distill-Qwen model using six completely different datasets. These six datasets included: training code with comments and without comments, training code containing 3D models and without 3D models, and training code with annotated and without annotated. The results indicated that when comments were added to the training code, the model's performance metrics improved across the board, with significant increases in APR and pass@k. This demonstrates that annotating the code helps the model enhance its ability to reason about context in long texts.Then, since the functions for drawing 2D and 3D sketches using the ezdxf library are entirely different and require different import libraries, we trained the model separately on a pure 2D dataset and a dataset mixed with 3D models to evaluate the impact of dataset content on model performance. The results showed that although the model's ACC-F and ACC-P experienced a certain degree of decline after mixing in 3D data, ACC-G improved. This is because the code for 3D models in the dataset is generally more concise compared to annotated 2D sketches; as long as the code compiles successfully, it yields nearly correct results. Since ACC-G is defined as the ratio of correctly outputted graphics to the total number of successful compilations, the inclusion of 3D data actually led to an increase in ACC-G. Therefore, we conclude that the impact of 3D data on the model is not significant, and CAD-Coder possesses generation capabilities for both 2D and 3D.Finally, we compared the performance metrics of the model on a purely unannotated dataset (including unannotated 2D sketches and 3D models) with those after incorporating an annotated dataset. The results indicated that CAD-Coder performed stably and well on the unannotated dataset. However, after adding annotations, the complexity of the training code increased, resulting in a decline in the model's performance capabilities.

In the final set of ablation experiments, we selected six large models as baseline models to test the generation performance of the CAD-Coder model under different baseline conditions. The primary baselines were the original Llama and Qwen models, followed by their respective large models focused on code generation, and finally the distilled versions of these two models. The experimental results indicated that the distilled model exhibited the best generation capabilities, followed by the baseline models, while the models specifically focused on code generation performed the worst. The reason for this is that the distilled model endows CAD-Coder with stronger reasoning abilities, whereas the models dedicated to code generation, due to certain prior knowledge, tend to generate other types of code that can disrupt the experimental results.

In the second set of ablation experiments, we conducted full parameter fine-tuning and LoRA fine-tuning on the distilled model to compare the effects of different methods on the model's generation capabilities. The results indicate that the LoRA method is more suitable for our task, as our dataset is not very large, and the LoRA method helps reduce the occurrence of overfitting.


\section{Details of Size Constraints} \label{sec:Details of Size Constraints}

In this section, we present the details of the size constraints used in our method.

\noindent\textbf{a.} Tangent circle constraint (circumtangent circle system)
\[ {d_{center} = R_{major} + R_{minor}} \]
\( R_{major} \): Radius of main structure circle (base circle)  
\( R_{minor} \): Dependent circle radius (constrained circle)

\noindent\textbf{b. }Hexagon nut opposite side width constraint
\[ {S_{hex} \geq 1.5d_{nominal} + 0.2d_{internal}} \]  
\( S_{hex} \): Width of opposite side of nut

\noindent\( d_{nominal} \): Nominal diameter  

\noindent\( d_{internal} \): Inner diameter of thread 

\noindent\textbf{c.} Flange bolt hole distribution circle diameter
\[ D_{pcd} \geq D_{bore} + 2.5D_{bolt} \]
\( D_{bore} \): Pipe aperture

\noindent\( D_{bolt} \): Nominal diameter of bolt

\noindent\textbf{d. }Ball quantity and size constraints of rolling bearings
\[ n_{ball} = \left\lfloor \frac{\pi(D_{outer} - D_{inner})}{2.2d_{ball}} \right\rfloor \]  
\(D_{outer}\): Inner diameter of outer ring

\noindent\(D_{inner}\): Outer diameter of inner ring

\noindent\( n_{ball}\): Quantity of ball

\noindent\(d_{ball}\): Diameter of ball

\noindent\textbf{e. }Involute gear tooth root transition curve constraints
\[ \rho_{root} \geq 0.25m_n \]
\( m_n \): The normal modulus

\section{Parent Code Example} \label{sec:Parent Code Example}
\begin{figure*}[hbp] 
  \centering
    \scalebox{1.0}{
\includegraphics[width=\textwidth,height=\textheight,keepaspectratio]{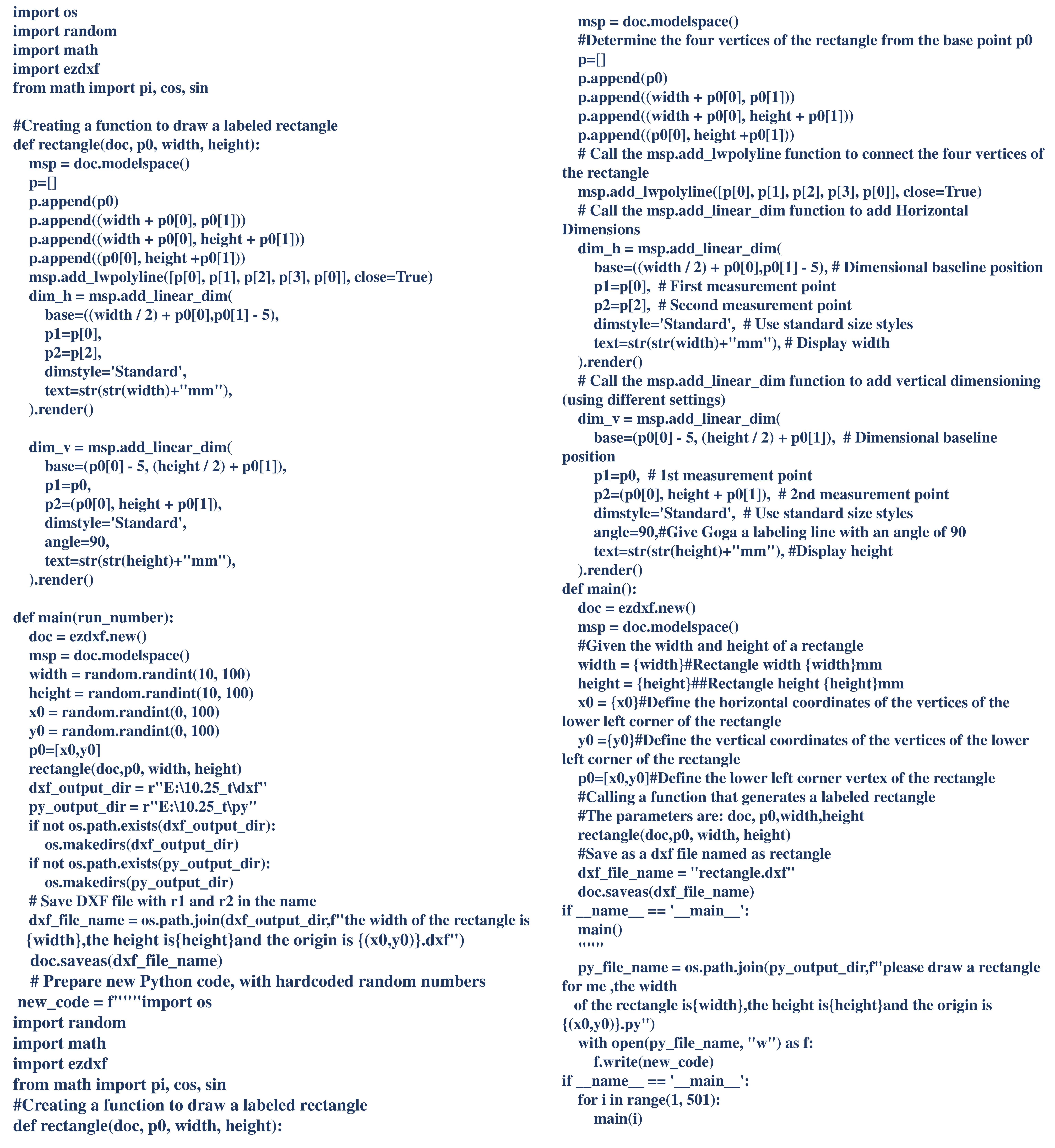}
}
  \caption{Annotated rectangle generation parent code.}
  \label{code}
\end{figure*}

As shown in the Figure\textbf{~\ref{code}}, it displays the parent code for generating annotated rectangle script code in our dataset. 

\end{document}